\documentclass[aps,prd,amssymb,eqsecnum,nofootinbib,floatfix, a4paper,twocolumn]{revtex4}

\usepackage{mathrsfs}
\usepackage{latexsym,amsmath,amsfonts,amssymb}
\usepackage{bm}

\usepackage{graphicx}

\usepackage{xcolor}  

\allowdisplaybreaks

\newcommand{\be}{\begin{equation}}
\newcommand{\ee}{\end{equation}}
\newcommand{\bea}{\begin{eqnarray}}
\newcommand{\eea}{\end{eqnarray}}

\newcommand{\sch}{{Schwarzschild \,}}
\newcommand{\cY}{{\cal Y}}

\def\k{\kappa}
\def\rs{r_{\star}}

\def\lam{{\lambda}}

\def\d{\partial}
\def\l{\left(}
\def\r{\right)}

\def\t0{\tilde{0}}

\def\c_F{{\cal F}}
\def\c_R{{\cal R}}

\newcommand{\bg}{\begin{gather}}
\newcommand{\eg}{\end{gather}}
\newcommand{\bseq}{\begin{subequations}}
\newcommand{\eseq}{\end{subequations}}

\begin{document}

\title{Perturbations of a Schwarzschild black hole in torsion bigravity 
}

\author{Vasilisa \surname{Nikiforova}}
 
\affiliation{Institut des Hautes Etudes Scientifiques, 
91440 Bures-sur-Yvette, France}

\date{\today}

\begin{abstract}
In this paper we pursue the study of linear perturbations around a Schwarzschild black hole in a generalized Einstein-Cartan theory of gravity, called torsion bigravity. This theory contains both massless and massive spin-2 excitations.  Here we consider non spherically-symmetric perturbations with generic multipolarity $L \geq 1$. We extend the conclusion of linear stability, previously obtained for $L=0$ [Phys. Rev. D {\bf 104}, 024032], to the generic $L \geq 1$ case. We prove that the mass $\kappa$ of the massive spin-2 excitation must be large enough, namely $\kappa r_h > \sqrt{1+\eta}$, to avoid the presence of singularities in the perturbation equations. The perturbation equations are shown to have a triangular structure, where massive spin-2 excitations satisfy decoupled equations, while the Einstein-like massless spin-2 ones satisfy inhomogeneous equations sourced by the massive spin-2 sector. We study quasi-bound states, and exhibit some explicit complex quasi-bound frequencies. We briefly discuss the issue of superradiance instabilities. 
\end{abstract}

\maketitle

\section{Introduction and reminder of torsion bigravity} \label{sec1}
Torsion bigravity is a modified theory of gravity which contains two independent dynamical fields: a space-time metric $g_{\mu\nu}$ (of signature mostly plus) and a metric-compatible ($\nabla ^{(A)}g=0$) affine connection ${A^\lam}_{\mu\nu}$ with torsion ${T^\lam}_{[\mu\nu]}$. The dynamics of these two fields is described by the following Lagrangian density:
\bea \label{lag0}
L&=& \frac{\lambda}{1+\eta} R[g]+  \frac{\eta\lambda}{1+\eta} F[g, A] \nonumber \\
&&+ \frac{\eta\lambda}{\k^2}\left( F_{(\mu \nu)}[A] F^{(\mu \nu)}[A] - \frac13 F^2[g, A] \right) \nonumber \\
&& + c_{34}F_{[\mu \nu]}[A]F^{[\mu \nu]}[A] \,.
\eea
Here $R[g]$ is the scalar (Riemannian) curvature of $g_{\mu\nu}$, while $F_{\mu \nu}[A] \equiv {F^\lam}_{\mu\lam\nu}[A]$ is the Ricci tensor of the torsionfull connection ${A^{\lambda}}_{\mu\nu}$, with corresponding scalar curvature  $F[g, A] \equiv g^{\mu\nu}F_{\mu\nu}[A]$. In Eq.~\eqref{lag0}, $F_{(\mu \nu)}$ and $F_{[\mu \nu]}$ respectively denote the symmetric and antisymmetric parts of $F_{\mu \nu}$. 

The spectrum of linear perturbations, on all the torsion bigravity backgrounds investigated until now, consists of a massless spin-2 perturbation and a massive spin-2 one of mass (or inverse range) $\k$. The massive spin-2 perturbation is contained in the perturbation of the torsion. 
The parameter $\lambda=\frac{1}{16\pi G_0}$ measures the gravitational coupling of the massless spin-2 field. The (dimensionless) parameter $\eta$ measures the ratio between the couplings of the massive and of the massless spin-2 fields. The $c_{34}$ parameter drops out of the linear consideration of the perturbations of torsionless backgrounds, as will be the case here.

The \sch spacetime is an exact solution of torsion bigravity\footnote{It is also the only possible generic asymptotically flat spherically-symmetric black hole solution \cite{Nikiforova:2020sac}. However, in the limit $\k \to 0$ there exist asymptotically flat black hole solutions \cite{Nikiforova:2020oyp}.} (see Ref.~\cite{Nikiforova:2009qr}). Answers for questions of stability and black hole observables find themselves in the investigation of  perturbations around \sch background. We started this investigation in a previous paper \cite{Nikiforova:2021xcj}, where we studied spherically symmetric perturbations around a \sch solution. In the present paper, we tackle the non-spherically symmetric perturbations. To this end, we decompose perturbations in spherical tensorial harmonics \`a la Regge-Wheeler-Zerilli (see \cite{Regge:1957td, Zerilli:1970se}). Let us recall that perturbations are separated in odd and even sectors, and are mainly characterized by their total angular momentum $L$, so that the spherically-symmetric case \cite{Nikiforova:2021xcj} corresponds to $L=0$.

Let us highlight the main results of the present paper.

\smallskip
I. The first result  is that the non-spherically symmetric perturbations are described by the expected number of degrees of freedom. Namely, there are five degrees of freedom describing the massive spin-2 sector (and two additional degrees of freedom in the massless spin-2 sector). This can be seen from counting the number of arbitrary initial data in the even and odd sectors for a general value of $L$, see subsections \ref{subOddLgen} and \ref{subEvenLgen}. This result confirms the good behavior of torsion bigravity in having the same number of degrees of freedom as {\it ghost-free} bimetric gravity \cite{deRham:2010kj, Hassan:2011zd}. We recall that torsion bigravity was found to have the correct number of degrees of freedom around torsionless Einstein backgrounds \cite{Nikiforova:2009qr}, in spherically-symmetric star-like solutions \cite{Damour:2019oru}, and also around \sch black hole solutions, as shown in \cite{Nikiforova:2021xcj} and here.

II. A second result concerns constraints on the mass $\k$ of the massive spin-2 excitation. When studying spherically-symmetric perturbations in \cite{Nikiforova:2021xcj}, we found that, in order to avoid having singularities in the equations which describe spherically-symmetric perturbations, the following constraint has to be necessarily imposed:
\be
\k r_h > \sqrt{1+\eta} \label{c} \,,
\ee
where $r_h$ is the \sch radius of the perturbed black hole. Which means that it is impossible to meaningfully discuss spherically-symmetric perturbations if this constraint is not satisfied. Here we prove that the need for this inequality extends to the case of non-spherically symmetric perturbations. Namely, the satisfaction of the constraint \eqref{c} is necessary and sufficient to avoid vanishing denominators in the equations describing the dynamics of generic perturbations. 

III. The third  result of the present paper concerns the existence of quasi-bound states, and the issue of stability. A quasi-bound state is a configuration of linear perturbations in the form of a Gamov-like process with a wave escaping towards the horizon, while it decays at infinity. The quasi-bound states are wave representations of particles turning around black holes and falling eventually on the black hole. These wave processes are detectors of instabilities: depending on the sign of the imaginary part of the frequency of a quasi-bound state, it can be stable (decaying in time) or unstable (growing in time). The presence of an unstable quasi-bound state would indicate the transformation of a black hole into some other object (in ghost-free bimetric gravity, \sch black holes are expected to decay into hairy black holes \cite{Brito:2013xaa} due to the presence of unstable quasi-bound states \cite{Babichev:2013una, Brito:2013wya}).

In  Ref.~\cite{Nikiforova:2021xcj} we proved that the spherically-symmetric perturbations, $L=0$, are stable when the constraint \eqref{c} is imposed. The aim of the present paper is to extend this result to the case of non-spherically symmetric perturbations, $L>0$. To this end, first, we prove analytically the stability of the $L=1$ odd sector. Then (following the strategy used in ghost-free bimetric gravity \cite{Brito:2013wya}), we give arguments for the stability of all modes in all sectors. Namely, we studied quasi-bound states in the even dipole ($L=1$) and even multipole ($L>1$) sectors. For each discovered quasi-bound state, the imaginary part of the frequency was found to have the correct (negative) sign.

IV. All said before concerns the massive sector of perturbations. In addition, we shall prove that the stability of the massive spin-2 perturbations is sufficient for establishing the stability of the massless sector, too. The reason is that, as we shall show, the equation describing the dynamics of the massless spin-2 sector is simply the usual Einstein equation with a non-trivial right-hand side depending on the massive spin-2 sector. In other words, a triangular decomposition of massless and massive components takes place. 
Thus, the absence of instabilities in the massive sector, and the usual Regge-Wheeler-Zerilli results concerning the stability of Einsteinian massless perturbations, e.g., \cite{Wald1979}, imply the absence of instabilities in the massless sector.

V. Though we mostly consider here perturbations around a \sch black hole, we should not forget about rotating black holes and about an important phenomenon which appears in the case of Kerr black hole, namely,  {\it superradiance} \cite{Zel'dovich1971}. Indeed, it was first shown in \cite{Damour:1976kh} (see also \cite{Zouros:1979iw, Detweiler:1980uk}) that in some cases the superradiance phenomenon implies the existence of exponentially unstable quasi-bound-states for massive particles in a rotating black hole background. As a consequence, the observation of rotating black holes (thus, the absence of such superradiance-induced instabilities) gives bounds to the mass of massive particles, e.g., \cite{Brito:2020lup, Stott:2020gjj}. 

Using results on superradiance instabilities for usual massive spin-2 excitations \cite{Brito:2020lup, Stott:2020gjj}, combined with  the WKB study of superradiance instabilities of massive scalar particles around a rotating black hole \cite{Zouros:1979iw}, we will obtain the following upper bound on the range of the dynamical torsion in torsion bigravity:
\be
\k^{-1} \lesssim  20 \; {\rm km} \;.
\ee
Let us note that this bound does not restrict the range of $\k$ more than the phenomenological bound obtained in \cite{Nikiforova:2021xcj} from applying the constraint \eqref{c} to an hypothetical $2M_{\odot}$ black hole, namely,
\be
\k^{-1} <  6 \; {\rm km} \;.
\ee
See Sec.~\ref{super} for more details. 

This paper is organized as follows. In Sec.~\ref{RWZsec} we explain the triangular decomposition of massless and massive perturbations, and write down the equation which describes the massive sector. In Sec.~\ref{equations} we obtain the systems of equations which describe the dynamics of even and odd (massive) perturbations, separately for the dipolar ($L=1$) and multipolar ($L>1$) cases. We count the number of degrees of freedom. 
In Sec.~\ref{search_key} we explain the approach we use to look for quasi-bound states. In Sec.~\ref{OddSec} we analytically prove the stability of the odd dipole perturbations, and exhibit some quasi-bound states. In Sec.~\ref{EvenDipStab} we exhibit some quasi-bound states of the even dipole perturbations, and discuss the issue of stability. In Sec.~\ref{EvenMulti} we exhibit some quasi-bound states of the even sector for $L>1$, and discuss the issue of stability.  
Finally, in Sec.~\ref{super} we discuss superradiant instabilities in torsion bigravity.


\section{Separating the dynamics of massive and massless spin-2 perturbations} \label{RWZsec}
As mentioned above,  Ref.~\cite{Nikiforova:2009qr} showed that a \sch spacetime is an exact solution of torsion bigravity. The evolution of the linear {\it massive} spin-2 perturbations above any torsionless Einstein background (the \sch spacetime being a particular case of an Einstein spacetime), was found in Ref.~\cite{Nikiforova:2009qr} to be described by a separate equation (see Eq.~(34) there\footnote{Note that the Eq.~(34) was obtained in \cite{Nikiforova:2009qr} for a more general theory which contained also a massive spin-0 excitation.}). This equation can be written in our notations as follows:
\bea
&& \nabla^2u_{\mu\nu} - \nabla^{\rho}\nabla_{\mu}u_{\rho\nu} - \nabla^{\rho}\nabla_{\nu}u_{\rho\mu} +\nabla_{\mu}\nabla_{\nu}u \nonumber \\
&&+ g_{\mu\nu}\l \nabla^{\rho}\nabla^{\sigma}u_{\rho\sigma} - \nabla^2 u \r - \k^2\l u_{\mu\nu} - g_{\mu\nu}u \r\nonumber \\
&& - (1+\eta)u^{\rho\sigma}W_{\mu\rho\nu\sigma}=0\,, \label{eqTor}
\eea
where 
\be
u_{\mu\nu}=F_{(1)\mu\nu} - \frac{1}{6}g_{\mu\nu}F_{(1)}\,, \label{eqGrav}
\ee
and $F_{(1)\mu\nu}$ is the first-order perturbation of the Ricci tensor $F_{\mu\nu}[A]$ of the torsionfull connection $A$. Here and below $\nabla_{\mu}$ denotes the Riemannian covariant derivative written through the Christoffel symbols of the considered background metric  $g_{\mu\nu}$.

Note that in the formal limit $\eta \to -1$ (which is not physically allowed in torsion bigravity where $\eta>0$) Eq.~\eqref{eqTor} reduces to the usual Fierz-Pauli-like equation satisfied in ghost-free bimetric gravity.

Once Eq.~\eqref{eqTor} is solved, the dynamics of the {\it massless} spin-2 sector of torsion bigravity is then described by rewriting Eq.~\eqref{eqGrav}. Indeed, the latter can be rewritten as 
\be
F_{(1)\mu\nu}=u_{\mu\nu}+\frac{1}{2}g_{\mu\nu}u \, ,\label{Fandu}
\ee
where the right-hand side will henceforth be considered as a source for the perturbed torsionfull Ricci-tensor 
entering the left-hand side. Indeed,
the torsionfull Ricci tensor $F_{\mu\nu}[A]$ can be decomposed into the usual Ricci tensor of General Relativity (GR) $R_{\mu\nu}[g]$ constructed from the Christoffel connection, and contorsion terms, as follows: 
\bea
F_{\mu\nu}&=&R_{\mu\nu} +  \nabla_{\lambda}{K^\lambda}_{\mu\nu} - \nabla_{\nu}{K^\lambda}_{\mu\lambda} \nonumber \\
&& + {K^\lambda}_{\sigma\lambda}{K^\sigma}_{\mu\nu} -  {K^\lambda}_{\sigma\nu}{K^\sigma}_{\mu\lambda}  \,.
\eea
Thus, for the first-order perturbation around a torsionless background we have
\be
F_{(1)\mu\nu}=R_{(1)\mu\nu} +  \nabla_{\lambda}{{K_{(1)}}^\lambda}_{\mu\nu} - \nabla_{\nu}{{K_{(1)}}^\lambda}_{\mu\lambda} \;, \label{Fmunu1}
\ee
where $R_{(1)\mu\nu}$ is the linear perturbation of the Ricci tensor of the perturbed metric $g_{\mu\nu} = g_{(0) \mu\nu} + g_{(1) \mu\nu}$.
The contorsion tensor is related to the torsion by
\be
K_{\mu\nu\rho} =\frac{1}{2}\l T_{\mu\nu\rho} +T_{\nu\rho\mu} - T_{\rho\mu\nu} \r \,, \label{KThroughT}
\ee 
while the perturbation of the torsion tensor can be expressed through $\nabla_{\rho}F_{(1)\mu\nu}$ according to Eqs.~(26), (30), (31) of \cite{Nikiforova:2009qr}. Namely, omitting indices, one can schematically write that
\be
T_{(1)\mu\nu\rho} \sim \frac{1}{\k^2} \l 1+\frac{W}{\k^2}\r \nabla F_{(1)}  \,,
\label{TThroughU}
\ee
where $W$ is the background Weyl tensor. Substituting Eq.~\eqref{TThroughU} into Eq.~\eqref{KThroughT}, and then substituting the result into Eq.~\eqref{Fmunu1}, one obtains that $F_{(1)\mu\nu}$ can be decomposed into the sum of the GR Ricci tensor $R_{(1)\mu\nu}$ and of some derivatives of $u_{\mu\nu}$, in the following symbolic manner:
\be
F_{(1)\mu\nu} \sim R_{(1)\mu\nu} + \left\{ \nabla \left[ \l 1+\frac{W}{\k^2} \r \frac{\nabla F_{(1)}}{\k^2} \right]   \right\}_{\mu\nu}\,.
\ee
This means that the Ricci tensor $R_{(1)\mu\nu}$ can be reexpressed as
\be
R_{(1)\mu\nu} \sim F_{(1)\mu\nu} + \frac{1}{\k^2} \left\{  \nabla \left[ \l 1+\frac{W}{\k^2} \r \nabla  F_{(1)} \right] \right\}_{\mu\nu} \,.
\ee
Taking into account the relation between $F_{(1)\mu\nu}$ and $u_{\mu\nu}$, Eq.~\eqref{Fandu}, we can write that
\be
R_{(1)\mu\nu} \sim u_{\mu\nu} + \frac{1}{\k^2} \left[  \nabla\nabla u + \nabla\l \frac{W}{\k^2} \nabla u \r   \right]_{\mu\nu} \,.
\ee
Thus, the dynamics of the massless spin-2 sector is described by the following inhomogeneous linearized Einstein's equation
\be
R_{(1)\mu\nu} - \frac{1}{2}R_{(1)}g_{\mu\nu}=T_{\mu\nu}^{\rm eff}  \,,  \label{EqEinstEf}
\ee
where the left-hand side is $\sim \Box g_{(1)\mu\nu}$, while the right-hand side is an effective stress-energy tensor of the form
\be \label{TmunuEf}
T_{\mu\nu}^{\rm eff} \sim u_{\mu\nu} + \frac{1}{\k^2} \left[  \nabla\nabla u + \nabla\l \frac{W}{\k^2} \nabla u \r   \right]_{\mu\nu} \,. 
\ee
The system, Eqs.~\eqref{EqEinstEf}--\eqref{TmunuEf}, is a well-posed computational problem, which we plan to address in future work. One can see that the perturbations of the metric depend on the massive spin-2 perturbations. Since the perturbations of metric can be directly detected on the Earth by various gravitational-wave detectors, they deserve a detailed study. 

Thus, the stability issue of the metric perturbations completely depends on the stability issue of the massive sector of perturbations. The vacuum perturbations of a \sch black hole are known to be linearly stable \cite{Wald1979}, and if, in addition, the right-hand side of Eq.~\eqref{EqEinstEf} is also stable, one will have no instabilities in the solutions of Eq.~\eqref{EqEinstEf}.

\section{Evolution of massive spin-2 perturbations and Regge-Wheeler-Zerilli decomposition} \label{equations}
As is usual for perturbations of spherically-symmetric black holes, we decompose massive spin-2 perturbations, as described by the symmetric tensor $u_{\mu\nu}$, both in frequency (factor $e^{-i\omega t}$) and in tensorial spherical harmonics \`a la Regge-Wheeler-Zerilli \cite{Regge:1957td, Zerilli:1970se}. This decomposition introduces the total angular momentum $L$ and separates perturbations in odd and even sectors.
As to the projection $M$ of the momentum $L$ on the $z$-axis, we follow Regge and Wheeler who emphasized that, due to the spherical symmetry of the background, for a given $L$, all the values of $M$ lead to the same radial equation. Thus, one can take $M=0$. This simplifies the angular dependence and allows one to work with Legendre polynomials $P_L(\theta)$ instead of working with complete spherical harmonics $Y_{LM}(\theta, \phi)$. We invite the reader to look at Appendix \ref{apA} for notations and details.

After inserting $u_{\mu\nu}$ decomposed in both frequency in spherical harmonics, both the time- and angular dependences can be factored out, and one is left with pure radial equations of evolution for the radial field variables. 

The equations describing the perturbations, as well as the number of degrees of freedom, is different for $L=1$ and $L>1$. So we will consider the case $L=1$ separately.
\subsection{Odd sector, $L>1$: two degrees of freedom}\label{subOddLgen}
As is clear from the tensorial spherical harmonics decomposition of Appendix \ref{apA}, there are 3 variables describing the odd-parity perturbation $u_{\mu\nu}$ for generic $L>1$: $h_0$, $h_1$ and $h_2$. After substituting the decomposition \eqref{DecOdd} into the general equation \eqref{eqTor}, factoring out the angular part and time dependence, we obtain the following three radial differential equations ($h^{'} \equiv d h/d r$):
\bea
&&\left[2 L(L+1) r  + 2 \k^2 r^3  + r_h (\eta-3)\right] h_0  \nonumber \\
 &&\quad - 
 4 i r (r - r_h) \omega  h_1 - 
 i (L-1) (2 + L) r \omega  h_2  \nonumber \\ 
  &&\quad - 
 2 i r^2 (r - r_h) \omega  h_1^{'}- 2 r^2 (r - r_h) h_0^{''} =0 \,, \label{OddL>1-1}
 \\
 &&  \left[ 2 (L-1)(2+L) r^2 - 2 \k^2 r^3 r_h - r_h^2 (1 + \eta) \right. \nonumber \\
 && \quad \left. + 
 r r_h (5 - 2 L - 2 L^2 + \eta) + 2 r^4 (\k^2 - \omega^2) \right] h_1  \nonumber \\
    && \quad -4 i r^3 \omega  h_0  - 
 2 (L-1) (2 + L) (r - r_h) h_2 \nonumber \\
 && \quad + 
 2 i r^4 \omega  h_0^{'} + (L-1) (2 + L) r (r - r_h) h_2^{'} =0 \,, \quad \label{OddL>1-2}
   \\
&& 2 i r^4 \omega  h_0 + 
 2 r (r - r_h) r_h h_1+ 
 2 r^2 (r - r_h)^2 h_1^{'} \nonumber \\
 && \quad  + \l \right.2 r^2 - \k^2 r^4 - 5 r r_h + \k^2 r^3 r_h + 3 r_h^2 + 
    r r_h \eta  \nonumber \\
    && \left. \quad - r_h^2 \eta + r^4 \omega ^2 \r h_2  - 
 r (2 r - 3 r_h) (r - r_h) h_2^{'} \nonumber \\
 && \quad + 
 r^2 (r - r_h)^2 h_2^{''} =0 \,. \label{OddL>1-3}
\eea

The combination  
\bea && i r^3 \omega\,{\rm Eq.}~\eqref{OddL>1-1}  + (r-r_h) \left[r\, {\rm Eq.}~\eqref{OddL>1-2}^{'} -2\,{\rm Eq.}~\eqref{OddL>1-2} \right] \nonumber \\
&& - (L-1)(L+2)\, {\rm Eq.}~\eqref{OddL>1-3}   \nonumber 
\eea
gives an equation which is algebraic in $h_0,h_1,h_2$ and $h_1^{'}$. We use it to express $h_0$ in terms of $h_1,h_2$ and $h_1^{'}$ as
\bea
h_0&=& \frac{1}{r^3 \left[2 \k^2 r^3 + r_h(1 + \eta) \right] \omega }
 i (r - r_h) \left\{\left[2 k^2 r^3 (2 r - r_h) \right. \right.\nonumber \\
 && \left. - (r - 2 r_h) r_h (1 + \eta )\right] h_1 \nonumber \\
 && + (L-1)(L+2) \left[\k^2 r^3 - r_h (1 + \eta )\right] h_2 \nonumber \\
 &&\left.+ 
    r (r - r_h) \left[2 \k^2 r^3 + r_h(1 + \eta) \right] h_1^{'} \right\} \,.
\eea 
Substituting this expression in Eqs.~\eqref{OddL>1-2} and \eqref{OddL>1-3} we get a system of two {\it second-order} equations for $h_1$ and $h_2$. Introducing for convenience the new variables 
\be \label{Q1}
 Q \equiv f(r)h_1 \;, \qquad Z \equiv \frac{h_2}{2r} 
 \ee
where 
\be
f(r)\equiv 1-\frac{r_h}{r} \,,
\ee
and using the tortoise coordinate 
\be \label{tortoise}
r_\star \equiv r + r_h \ln{\l \frac{r}{r_h}-1\r} \;,
\ee
one can finally get the following system of two second-order equations for the two variables, $Q$ and $Z$:
\bea \label{sysOdd}
&& \frac{d^2}{d r_{\star}^2}Z(r) +\left(\omega^2 -f(r)V_Z[\k,\eta,L]\right)Z(r)=S_Z \,, \;\\
&&\frac{d^2}{d r_{\star}^2}Q(r) -f(r)\frac{3 r_h (1 + \eta)}{r \left[2 \k^2 r^3 + r_h(1 + \eta)\right]} \frac{d}{d r_{\star}}Q(r) \nonumber \\
&&+ \left(\omega^2 - f(r)V_Q[\k,\eta,L]\right)Q(r) = S_Q\,.
\eea
Here
$$ \Lambda \equiv L(L+1) \,,$$
the terms $V_Z$ and $V_Q$ are as follows:
\bea \label{potOdd}
&&V_Z[\k,\eta,L] \equiv  \frac{1}{r^3 \left[2 \k^2 r^3 + r_h(1 +  \eta) \right]}\left\{2 \k^4 r^6  \right. \nonumber \\
&&  + \k^2 r^3 \left[2 (\Lambda-2) r + r_h (1- \eta) \right] \nonumber \\
&& - 
   r_h (1 + \eta ) \left[2 (\Lambda-2) r + r_h \eta \right]\left. \right\} \,, \\
&& V_Q[\k,\eta,L] \equiv \frac{1}{2 r^3 \left[2 \k^2 r^3 + r_h(1 + \eta)\right]^2} \left\{8 \k^6 r^9 \right. \nonumber \\
&& + 
  r_h^2 (1 + \eta )^2 \left[2 (\Lambda-5) r + r_h (9 + \eta )\right] \nonumber \\
  && + 
  4 \k^4 r^6 \left[2 (\Lambda + 4 ) r + r_h (3 \eta -13 )\right] \nonumber \\
  && \left. + 
  2 \k^2 r^3 r_h (1 + \eta ) \left[4 (\Lambda-5) r + r_h (13 + 3 \eta )\right] \right\} , \qquad
\eea
and the ``sources" $S_Z$ and $S_Q$ read as:
\bea \label{sourceOdd}
&&S_Z \equiv f(r) \frac{ 4 \k^2 r^3 - r_h (1 + \eta)}{r^2 \left[2 \k^2 r^3 + r_h( 1+ \eta)\right]} Q(r) \,, \\
&& S_Q \equiv f(r)\frac{3 (\Lambda-2) r_h (1 + \eta)}{r \left[2 \k^2 r^3 + r_h(1 + 
   \eta)\right]} \frac{d}{d r_{\star}}Z(r) \nonumber \\
   && + f(r) \frac{(\Lambda-2)}{r^3 \left[2 \k^2 r^3 + r_h (1+ \eta)\right]^2} \left\{ \right. 4 \k^4 r^6 (2 r - 3 r_h)\nonumber \\
   && - 
   2 \k^2 r^3 (14 r - 15 r_h) r_h (1 + \eta) \nonumber \\
   &&- (7 r - 
      9 r_h) r_h^2 (1 + \eta)^2 \left. \right\} Z(r) \,.
\eea
An interested reader can compare Eqs.~\eqref{sysOdd}--\eqref{sourceOdd} with equations (32)--(35) of \cite{Brito:2013wya} and see that, when $\eta=-1$, the first ones transform into the last ones. 

Thus, for $L>1$ in the odd sector one ends up with a system of {\it two second-order differential equations}, which corresponds to {\it two degrees of freedom}.

\subsection{Odd sector, $L=1$: one degree of freedom}
In the case $L=1$, the variable $h_2$ does not exist in the decomposition \eqref{DecOdd}, and the third equation of the previous section, Eq.\eqref{OddL>1-3}, does not apply. The system then contains one second-order equation and one first-order equation for $h_0$ and $h_1$:
\bea
&&\left[4 r + 2 \k^2 r^3 + r_h ( \eta-3)\right] h_0 - 
 4 i r (r - r_h) \omega  h_1 \nonumber \\
&& \quad  - 
 2 i r^2 (r - r_h) \omega  h_1^{'} - 
 2 r^2 (r - r_h) h_0^{''} = 0  \,,
 \\
 && \left[ 2 \k^2 r^3 (r - r_h) + r_h(r-r_h) (1 + \eta)  - 
    2 r^4 \omega ^2 \right] h_1 \nonumber \\
 &&  \quad -4 i r^3 \omega  h_0  + 2 i r^4 \omega  h_0^{'} =0 \,.
\eea
It is convenient to work with $h_1$. After doing some simple algebraic and differential manipulations with these two equations one can obtain a second-order differential equation describing the evolution of $h_1$.  
Introducing a new variable $Q$ (not to be confused with the one defined in Eq.~\eqref{Q1} above)
\be
Q \equiv f(r) \sqrt{1+\frac{(1+\eta)r_h}{2\k^2 r^3}} h_1 \,,
\ee
one can write down the equation describing the dynamics of the odd dipole perturbations, as the following equation 
\bea
&&f(r)^2 Q^{''}(r) + f(r)\frac{r_h}{r^2} Q^{'}(r) \nonumber \\
&&+ \left( \omega^2 - V_1[r,r_h,\k,\eta] \right) Q(r)=0 \;,  \label{EqQOddDipole}
\eea
where
\bea
&&V_1[r,r_h,\k,\eta] = \frac{(r - r_h)}{4 r^4 \left[2 \k^2 r^3 + r_h(1 + \eta)\right]^2}\left\{ 96 \k^4 r^7  \right. \nonumber \\
&& + 16 \k^6 r^9 - 104 \k^4 r^6 r_h + 3 r r_h^2 - 
 8 \k^2 r^3 r_h^2 - 3 r_h^3  \nonumber \\
 &&  + 24 \k^4 r^6 r_h \eta   + 6 r r_h^2 \eta  + 
 4 \k^2 r^3 r_h^2 \eta  - 4 r_h^3 \eta  + 3 r r_h^2 \eta ^2 \nonumber \\
 && + 
 12 \k^2 r^3 r_h^2 \eta ^2 + r_h^3 \eta ^2 + 2 r_h^3 \eta ^3  \left. \right\} \;. \label{PotV1}
\eea
In terms of the tortoise coordinate \eqref{tortoise}, Eq.~\eqref{EqQOddDipole} takes the following Schr\"{o}dinger-like form: 
\be \label{EqQOddDipoleStar}
\frac{d^2}{d r_{\star}^2}Q(r) + \left( \omega^2 - V_1[r,r_h,\k,\eta] \right) Q(r)=0 \;.
\ee
Let us note in passing that Eq.~\eqref{EqQOddDipoleStar}, together with the potential \eqref{PotV1}, in the limit $\eta \to -1$ reduces to Eq.~(36) of \cite{Brito:2013wya}.

Given a solution of Eq.~\eqref{EqQOddDipole}, then $h_0$ is obtained as
\bea
h_0&=&\frac{i (r - r_h) \left[8 \k^2 r^3 + r_h(1 + \eta)\right] Q}{
 \sqrt{2} \k^2 r^5 \l 2 +\frac{ r_h(1 + \eta)}{k^2 r^3}\r^{
  3/2} \omega} \nonumber \\
  &&+ \frac{i \sqrt{2} (r - r_h) Q^{'}}{
 r \l2 +\frac{ r_h(1 + \eta)}{k^2 r^3}\r^{
  1/2} \omega} \,. 
\eea

Thus, the odd-sector perturbations in the case $L=1$ are described by {\it one second-order differential equation}, which corresponds to {\it one degree of freedom}.

\subsection{Even sector, $L>1$: three degrees of freedom}\label{evendipsec} \label{subEvenLgen}
The even-parity sector of the massive spin-2 field $u_{\mu\nu}$ is described, according to Eq.~\eqref{DecEven} in Appendix \ref{apA}, by seven field variables: 
\be
G, \overline{h}_0, H_0, \overline{h}_1, H_1, H_2, K\,. \label{VarEven}
\ee
After substituting the decomposition \eqref{DecEven} into the general equation \eqref{eqTor}, making a frequency decomposition, and factoring out the angular part, we obtain seven differential equations on seven variables \eqref{VarEven} [These original equations are available as a MATHEMATICA notebook in the {\it Supplemental Material}]. 
To be more precise, three of these equations are first-order, and four are second-order ordinary differential equations (ODEs). 

Since the reduction of this system is more complicated than in the odd case, we found convenient to first reduce the original seven ODEs to a system of {\it first-order} ODEs, by introducing the following  four auxiliary variables,
\bea  \label{auxvar}
G_p(r) &\equiv & G^{\prime}(r) \,, \nonumber \\
 h_{0p}(r) &\equiv & \overline{h}_{0}^{\prime}(r) \,,  \nonumber \\
 H_{0p}(r) &\equiv & H_{0}^{\prime}(r) \,,  \nonumber \\
 K_{p}(r) &\equiv & K^{\prime}(r) \,.
\eea
This way, the original system of 7 (generally speaking, second-order) differential equations can be transformed into an equivalent system of 11 first-order ODEs,
\be
 A_{\alpha i}(r)\, {\cal Y}_i^{\prime}(r) + B_{\alpha i}(r)\, {\cY}_i(r) =0 \label{11eq} \,,
 \ee
for 11 variables
\bea  \label{11var}
&&\left[{\cal Y}_{i}(r)\right]_{i=1,\ldots,11} \equiv \nonumber \\
&& 
 \{G, \overline{h}_0, H_0, \overline{h}_1, H_1, H_2, K, G_p, h_{0p}, H_{0p}, K_p  \} \,.\quad
 \eea
Here the index $\alpha=1,\ldots, 11$ labels the 11 linearized field equations, while the index $i=1,\ldots,11$ labels the 11 frequency-space perturbed field variables \eqref{11var}. We use Einstein's summation convention on all repeated indices (here: $i=1,\ldots,11$). The system \eqref{11eq} can be written in matrix form as
\be
\label{11eqmat}
 A_{I}\, {\cal Y}_I^{\prime} + B_I\, {\cY}_I =0  \,,
\ee
where $A_I$ and $B_I$ are $11\times 11$ matrices, while $\cY$ is an 11-dimensional column vector.

This system contains a certain number of algebraic constraints relating the 11 variables \eqref{11var}. To find all these constraints, we use the Dirac method of primary, secondary etc. constraints.

The procedure is the following. First, we found that the rank of the system \eqref{11eqmat} is 8. This means that there exist $11-8=3$ (primary) algebraic constraints
\be
v_I\,B_I\,\cY_I=0 \,, 
\ee
where  $v_I$ is a left null-vector of the matrix $A_I$ such that $v_I \,A_I=0$. Some of these constraints can be trivial or dependent on the other ones. In our case, the three left null-vectors yield three  independent non-trivial (primary) constraints on the 11 variables $\cY$ which we solve for $G_p, H_{0p}, K_p$ in terms of the eight residual variables, 
\be \label{YII}
\cY_{II} \equiv \{ G, \overline{h}_0, H_0, \overline{h}_1, H_1, H_2, K, h_{0p} \} \,.
\ee
After substituting $G_p(\cY_{II}), H_{0p}(\cY_{II}), K_p(\cY_{II})$ in the initial system of equations \eqref{11eqmat}, we obtain a new (secondary) system
\be
\label{11eq2mat}
 A_{II}\, {\cal Y}_{II}^{\prime} + B_{II}\, {\cY}_{II} =0  \,,
\ee
where ${\cal Y}_{II}$ is the $8$-dimensional column vector \eqref{YII}, while $A_{II}$ and $B_{II}$ are $11\times8$ matrices.

Applying to the system \eqref{11eq2mat} the same approach as to the system \eqref{11eqmat}, we obtain one independent algebraic secondary constraint which we use to express 
\be \overline{h}_1(\cY_{III}) \label{2constr} \ee
 in terms of the seven residual variables,
\be \label{YIII}
\cY_{III}\equiv \{ G, \overline{h}_0, H_0, H_1, H_2, K, h_{0p} \} \,.
\ee 
After substituting \eqref{2constr} in the system \eqref{11eq2mat} we obtain a new (tertiary) system
\be
\label{11eq3mat}
 A_{III}\, {\cal Y}_{III}^{\prime} + B_{III}\, {\cY}_{III} =0  \,,
\ee
where ${\cal Y}_{III}$ is the $7$-dimensional column vector \eqref{YIII}, while $A_{II}$ and $B_{II}$ are $11\times7$ matrices.
Pursuing this strategy with the system \eqref{11eq3mat} we find one left null-vector of the matrix $A_{III}$, and then obtain one independent algebraic tertiary constraint which we use to express 
\be h_{0p}(\cY_{IV}) \label{3constr} \ee
 in terms of the six residual variables,
\be \label{YIV}
\cY_{IV}\equiv \{ G, \overline{h}_0, H_0, H_1, H_2, K \} \,.
\ee 
After substituting \eqref{3constr} in \eqref{11eq3mat} we obtain a system of the form 
\be
\label{11eq4mat}
 A_{IV}\, {\cal Y}_{IV}^{\prime} + B_{IV}\, {\cY}_{IV} =0  \,,
\ee
where ${\cal Y}_{IV}$ is the $6$-dimensional column vector \eqref{YIV}, while $A_{II}$ and $B_{II}$ are $11\times6$ matrices. 

There are no more constraints hidden in the system \eqref{11eq4mat}: we have checked that 5 of these 11 equations can be written as linear combinations of the remaining 6 equations. Thereby, as a final result, we obtain a system of {\it 6 first-order differential equations on the 6 variables} $\cY_{IV}\equiv \{ G, \overline{h}_0, H_0, H_1, H_2, K \}$.

Thus, the $L>1$ case of the even sector is described by 6 first-order ODEs
\be
V_6^{'}(r)=A_6[r, r_h,\k,\eta,L,\omega]V_6(r) \,, \label{sysvar6}
\ee
where
\be
V_6(r) \equiv \{G,\overline{h}_0,H_0,H_1,H_2,K\} \,,
\ee
and where the structure of the matrix $A_6[r, r_h,\k,\eta,L,\omega]$ reads
\be
A_{6\;ij}=(i \omega)^n \left[ \frac{a_{ij}[r, r_h,\k,\eta,L] + \omega^2\,b_{ij}[r, r_h,\k,\eta,L]}{d_{ij}[r, r_h,\k,\eta,L]} \right] \,, \label{A6el}
\ee
with $n=0 \text{ or } 1$.  The explicit forms of Eqs.~\eqref{sysvar6}, and of the matrix $A_{6}[r, r_h,\k,\eta,L,\omega]$ are available in the {\it Supplemental Material}.

The denominators $d_{ij}[r, r_h,\k,\eta,L]$ do not depend on $\omega$ and consist of several factors. Apart from the factors that do not become zero at any $r>r_h$, there are the following factors which can create singularities outside the horizon ($r>r_h$):
\be 
\left(r^3-\frac{r_h(1+\eta)}{\k^2} \right)\,, \label{r3}
\ee
and
\be
\left(r^6- \frac{r_h^2\eta(1+\eta)}{\k^4} \right)\,. \label{r6}
\ee
The denominator \eqref{r3} already appeared in the study of the spherically-symmetric perturbations of a \sch black hole \cite{Nikiforova:2021xcj}. This denominator was the origin of the constraint \eqref{c} that we had to admit in this theory to avoid having singularities in perturbation equations. We remind the reader that we had to admit
\be
\k r_h > \sqrt{1+\eta} \,. \label{constraintK}
\ee 
Thus, we {\it extend the result} \eqref{constraintK} to the generic case $L>0$: one needs to satisfy this constraint to avoid singularities appearing in the equations guiding the evolution of perturbations. In other words, {\it no additional constraints} appear in the generic case $L>0$ with respect to the case of spherically symmetric perturbations. We also recall that the presence of the denominator \eqref{r6} has no importance as soon as the condition \eqref{constraintK} is satisfied. Indeed, if the condition \eqref{constraintK} is satisfied, the last term in the brackets in Eq.~\eqref{r6} is
\be
\frac{r_h^2\eta(1+\eta)}{\k^4} < \frac{r_h^6 \eta}{1+\eta} < r_h^6 \,,
\ee
thereby the factor \eqref{r6} never becomes zero for any $r>r_h$.

Let us also note in passing that some denominators $d_{ij}$ in \eqref{A6el} contain a factor $(L-1)$, which reflects the fact that the system for $L=1$ is very different, see the next subsection.

Given a solution of the system \eqref{sysvar6}, the corresponding value of  the remaining variable $h_1$ can be found from an algebraic expression of the following type (whose explicit expression can be found in {\it Supplemental Material}):
\be
\overline{h}_1=M_6[r, r_h,\k,\eta,L,\omega] V_6(r) \,.
\ee
The elements of the matrix $M_6[r, r_h,\k,\eta,L,\omega]$ have the following form:
\be
M_{6\;ij}=\frac{N[r,r_h,\k,\eta, L,\omega]}{D[r,r_h,\k,\eta, L](r-r_h)^2} \,,
\ee
where $D[r,r_h,\k,\eta, L]$ never becomes zero for any physical values of $r, r_h,\k,\eta,L$.

Let us finally remark that, evidently, instead of working with the system of first-order differential equations \eqref{sysvar6}, one can use 3 of these equations to eliminate 3 first-derivatives, and then end up with a system of 3 {\it second-order} equations. This corresponds to {\it three degrees of freedom} in the even-sector of perturbations in the case $L>1$.

\subsection{Even sector, $L=1$: two degrees of freedom}
When $L=1$, the $G$ variable does not exist in the spherical harmonics decomposition, because $W_1(\theta)=0$ (see 
Appendix \ref{apA}). The system of equations describing the even perturbations for $L=1$ contains six differential 
equations (generally speaking, second-order) for six variables $\overline{h}_0, H_0, \overline{h}_1, H_1, H_2, K$. By analogy with the 
$L>1$ case, introducing the three auxiliary variables, 
\bea  \label{auxvarL1}
 h_{0p}(\omega,r) &\equiv & \overline{h}_{0}^{\prime}(\omega,r) \,,  \nonumber \\
 H_{0p}(\omega,r) &\equiv & H_{0}^{\prime}(\omega,r) \,,  \nonumber \\
 K_{p}(\omega,r) &\equiv & K^{\prime}(\omega,r) \,,
\eea
one transforms the original system into a system of $9$ {\it first-order} ODEs for 9 variables. Using the same approach as 
in the previous section (i.e., with primary, secondary and tertiary constraints) we end up with a system of {\it four 
first-order differential equations for four variables} $
V_4(r) \equiv \{\overline{h}_0,\overline{h}_1,H_1,K\}$:
\be
V_4^{'}(r)=A_4[r, r_h,\k,\eta,L,\omega]V_4(r) \,. \label{EqEvenL1}
\ee
The system \eqref{EqEvenL1} describes {\it 2 degrees of freedom}.

The explicit form of Eqs.~\eqref{EqEvenL1} is given in the {\it Supplemental Material}. Let us just indicate the structure of the matrix $A_4[r, r_h,\k,\eta,L,\omega]$, namely
\be
A_{4\;ij}=(i \omega)^n 
\left[ \frac{\overline{a}_{ij} + \overline{b}_{ij}\omega^2 +\overline{c}_{ij}\omega^4 }{\overline{d}_{ij}\left( \overline{e}_{ij} + \overline{f}_{ij}\omega^2 \right)} \right] \,,
\ee
where each of the coefficients $\overline{a}_{ij}, \overline{b}_{ij}, \dots, \overline{f}_{ij}$ depends on $r, r_h, \k, \eta, L$. Some of the denominators $\overline{d}_{ij}$ contain factors $(r-r_h), (r-r_h)^2$, the factors $\l \k^2 r^3 - (1+\eta)r_h \r$ and $\l \k^4 r^6 - r_h^2\eta(1+\eta) \r$ which are already discussed in the previous subsection, and other factors
 that do not become zero at any $r$. 
 The $\omega$-dependent factor $ \l \overline{e}_{ij} + \overline{f}_{ij}\omega^2  \r $ can become zero, but this zero is spurious and does not lead to any singularity in the solutions.

\section{Search for quasi-bound states, and stability issues : strategy} \label{search_key}
We wrote down all the equations describing the even and odd sectors, and now we are ready to proceed with the search of quasi-bound states in both sectors. 
In this section, we describe the general approach that we use for looking for quasi-bound states. 

As was said before, a quasi-bound state corresponds to a solution that exponentially decays at infinity and becomes an incoming wave near the horizon [See Fig.~\ref{Odddipolere} as an example]. Such configurations exist for discrete complex frequencies $\omega_{QBS}=\omega_{R}+ i \omega_{I}$. As mentioned in the Introduction, the notion of quasi-bound state is directly related to the stability issue. Indeed, restoring the time dependence $e^{-i\omega t}$ one sees that the evolution of a quasi-bound state depends on the sign of the imaginary part of $\omega_{QBS}$ :
\be
e^{-i\omega t} = e^{-i\omega_R t}e^{\omega_I t} \,. \label{QBSDecay}
\ee
If the sign of $\omega_I$ is negative, the state is (exponentially) decaying in time, which means that it is stable. Similarly to what was done in Ref. \cite{Brito:2013wya}, we will consider that finding quasi-bound states (QBSs) having only negative imaginary parts gives strong evidence for the stability of the black holes. 

The absolute value of $\omega_I$ defines the lifetime of the quasi-bound state. Here we will be interested in looking for QBSs for which the absolute value of $\omega_I$ is small, 
\be
|\omega_I| \ll |\omega_R| \,, \label{wi}
\ee
so that they are bound by the black hole over a reasonably long lifetime. In addition, the real part should satisfy the bound-state condition $|\omega_R|<\k$, see below. 

One can find a sought field configuration by posing boundary conditions on the horizon and at infinity. Namely, at infinity, the required solution should (exponentially) decay as $\sim \exp^{-\sqrt{\k^2-\omega^2}r}$. On the horizon, one should allow only ingoing waves to propagate, which means that the complete time-dependent solution should behave near the horizon proportionally to
\be
e^{-i \omega (\rs+t)} \,, \label{QBSnearhor}
\ee
where $r_{\star}$ is defined in Eq.~\eqref{tortoise}. The condition \eqref{QBSnearhor} implies that, near the horizon, the sought solution should depend on $r$ as
\bea
&&e^{-i \omega \rs} = e^{-i \omega \left[  r + r_h\ln{\l \frac{r}{r_h}-1  \r}  \right] }  \nonumber \\
&& = e^{-i\omega r}\times \l \frac{r-r_h}{r_h} \r^{-i\omega r_h} \sim (r-r_h)^{-i\omega r_h} \,.\; \label{QBSnearhor2}
\eea

In practice, the following approach to the search of the quasi-bound states is convenient. 

Let us  first discuss the case (applicable to odd dipolar perturbations) where the perturbation is described by only one field variable satisfying one (second-order) field equation [see Eq.~\eqref{EqQOddDipole}]. We  determined the general solutions, near the horizon, of this equation satisfying the boundary condition \eqref{QBSnearhor2}. Namely, we insert in the field equation \eqref{EqQOddDipole} the following ansatz to describe the near-horizon behavior :
\bea
&&Q(r) = (r-r_h)^{-i\omega r_h} \left[ a_0(r-r_h)^{n_0} + a_1(r-r_h)^{n_0+1} \right. \nonumber \\
&& \left.+ a_2(r-r_h)^{n_0+2} + a_3(r-r_h)^{n_0+3} + \cdots  \right] \,. \label{NHE}
\eea 
Here the power $n_0$ depends on the regularity properties of the field $Q$ on horizon ($n_0=0$ if $Q$ is regular on the horizon, which is the case of the odd dipole). Inserting this ansatz in the field equation we determine the coefficients $a_1, a_2, ...$ of the near-horizon expansion \eqref{NHE} in terms of $a_0$. The value of $a_0$ can be chosen arbitrarily, for instance, $a_0=1$.

The idea is then to integrate the field equation, e.g., \eqref{EqQOddDipole}, starting from the horizon, with the boundary condition \eqref{NHE} searching for values $\omega_{QBS}=\omega_{R\;QBS}+i\omega_{I\; QBS}$ such that the solution decays exponentially  at infinity. In other words, $\omega_{QBS}$ must be a complex solution of the complex equation $Q(r_{far},\omega_{QBS})=0$, where $r_{far}$ is some sufficiently large value of $r$. To find such complex solutions $\omega_{QBS}$, we used a two-step scheme. In the first step, we  take
advantage of the assumed smallness of $\omega_I$,  Eq.~\eqref{wi}, to explore the variation of the (real)
modulus $|Q(r_{far}, \omega_{R})|$ as a function of a purely real frequency $\omega=\omega_{R}$.
When using a suitable normalisation of $Q$  we indeed found that $|Q(r_{far}, \omega_{R})|$ exhibited clear minima at several points along the real $\omega_R$ axis (see Fig.\eqref{Odddipolemins} as an example of such a plot). We then looked for complex solutions  $\omega_{QBS}=\omega_{R\;QBS}+i\omega_{I\; QBS}$ of $Q(r_{far},\omega_{QBS})=0$ near those real minima. As $Q(r_{far},\omega)$ is an analytic function of $\omega$, we found convenient to look for these solutions by using Newton's iteration method. 

Now let us discuss the case where perturbations are described by several field variables satisfying a system of coupled linear differential equations. For simplicity, let us assume that there are two fields, $\phi_1$ and $\phi_2$, and two {\it second-order} differential equations. Imposing ingoing waves, we find that the near-horizon expansions for $\phi_1$ and $\phi_2$ contain two independent parameters, say, $a_0$ and $b_0$. Thus, for each value of $\omega$ there exist infinitely many different solutions corresponding to the different choices of $a_0$ and $b_0$. The problem then consists in finding  (complex) values of $\omega$ such that  there exist a choice of $a_0$ and $b_0$ for which the corresponding solution decays exponentially at infinity. 

Since the system of differential equations is linear, the sought exponentially decaying solution, say
\be
\widehat{\phi} \equiv (\widehat{\phi}_1,\widehat{\phi}_2) \,,
\ee
can always be presented as a linear combination of two "basic" solutions,
\be
\widehat{\phi} =  \Phi_{bas} \, B\,, 
\ee
where $B$ is a column vector, and where $\Phi_{bas}$  is the following $2 \times 2$ matrix
\be
\Phi_{bas} \equiv \begin{pmatrix} \phi_{1\;a} & \phi_{1\;b} \\  \phi_{2\;a} & \phi_{2\;b} \end{pmatrix} \,.
\ee
Here $(\phi_{1\;a}, \phi_{2\;a})$ is the solution obtained with initial conditions $a_0=1, b_0=0$, while the solution $(\phi_{1\;b}, \phi_{2\;b})$ corresponds to $a_0=0, b_0=1$. The condition of exponential decay at infinity $\widehat{\phi}(r_{far}, \omega_{QBS})=0=\Phi_{bas}(r_{far}, \omega_{QBS}) B$ can be then satisfied only if the determinant of the matrix $\Phi_{bas}(r_{far}, \omega_{QBS})$ is zero. Thus, the following equation
\be
{\rm det}\left[ \Phi_{bas}(r_{far}, \omega_{QBS}) \right] =0 \,,
\ee
becomes a complex equation to find a complex solution $\omega_{QBS}$. Moreover, the determinant ${\rm det}\left[ \Phi_{bas}(r_{far}, \omega) \right]$ is a (complex) analytic function of $\omega$. We then apply the two-step procedure described above, to the function ${\rm det}\left[ \Phi_{bas}(r_{far}, \omega) \right]$.

In the next sections, we  apply this approach to find some quasi-bound states. We start with the dipole perturbations, Sec.~\ref{OddSec} and \ref{EvenDipStab}.

\section{Quasi-bound states and stability of the odd dipole sector} \label{OddSec}
\subsection{Stability}
To understand the stability issue of the odd sector in the dipolar case $L=1$, let us rewrite the equation \eqref{EqQOddDipoleStar} reintroducing the time-dependence. We get
\be
-\frac{\d^2}{\d t^2}Q + \frac{\d^2}{\d r_{\star}^2}Q - V_1 Q=0 \,. \label{Z-l}
\ee
Such an equation of motion can be obtained from the following Lagrangian
\be
L=\frac{1}{2}\l\frac{\d}{\d t}Q\r^2 - \frac{1}{2}\l\frac{\d}{\d r_{\star}}Q\r^2 - \frac{1}{2}V_1 Q^2 \,.
\ee
This Lagrangian leads to the following conserved energy:
\be
E = \frac{1}{2}\l\frac{\d}{\d t}Q\r^2 + \frac{1}{2}\l\frac{\d}{\d r_{\star}}Q\r^2 + \frac{1}{2}V_1 Q^2 \,. \label{energy}
\ee
If the potential $V_1$ is  positive everywhere outside the horizon, all the three terms in the expression \eqref{energy} are positive and, thus, any instability is impossible.
One can easily show that this is the case for the odd-dipole potential $V_1$ derived above. More precisely, if the condition \eqref{constraintK} is satisfied (and if $\eta$ is positive),
 the potential $V_1$ (see Eq.~\eqref{PotV1}) is positive for $r>r_h$. This fact proves the {\it stability} of the dipolar perturbations in the odd sector. 

Let us recall that in the case of a Zerilli-like equation that described spherically-symmetric perturbations ($L=$0) the potential was not always positive. In that case, stability was proven by using a bound-state counting result, see \cite{Nikiforova:2021xcj}. 

\subsection{Quasi-bound states} 
The potential $V_1(r)$ (see Eq.~\eqref{PotV1}) starts from the value $V_1(r=r_h)=0$ on the horizon, and then  asymptotically reaches the value $V_1(r\to\infty)=\k^2$ (see Fig.\ref{Odddipoleenergy}). For $\eta$ close enough to zero and $\k$ close enough to its starting constrained value $\k_{\rm min}=\sqrt{1+\eta}$, there exists a local minimum of the potential at $r/r_h \sim 8$.  This minimum disappears with increasing values of $\eta$ or/and $\k$. Though in general, the potential $V_1$ has no such local minimum, quasi-bound states continue to exist even in the absence of such a local minimum. While quasi-bound states are usually linked to  a minimum of a potential (corresponding
to a classical stable circular orbit), here they generically correspond to classical motions describing unstable circular orbits where a massive particle slowly inspirals towards the horizon.

Note that at $r \to \infty$ (which corresponds to the Newtonian limit), the far-field expansion of the potential \eqref{PotV1} reads
\be
V_1|_{r\to\infty} = \l 1-\frac{r_h}{r} \r \l  \k^2 + \frac{6}{r^2} + O\l\frac{1}{r^3}\r \r \,.
\ee
This corresponds to a usual large-$r$ Zerilli-like potential, 
\be
V_1|_{r\to\infty} =\l 1-\frac{r_h}{r} \r \l  \k^2 + \frac{l(l+1)}{r^2} + O\l\frac{1}{r^3}\r  \r \,,
\ee
with the value $l=2$ for the {\it orbital} angular momentum of a wave/particle. The last result is compatible with the quantum triangle rule of addition of angular momenta,
\be
|L-s| \leq l \leq L+s \,,
\ee
where $L$ is the total angular momentum, $l$ is the orbital angular momentum, and $s$ is the spin of the perturbation. [Here, we have $L=1$, $s=2$, and $l=2$.]

To investigate an example of quasi-bound state, we took particular values of the parameters. First, we choose units where $r_h=1$. Next, according to the phenomenological limit (10.8) of \cite{Damour:2019oru}, for $\k^{-1} \lesssim 10\; {\rm km}$ the physically relevant values of $\eta$ are $\eta \lesssim 3\times 10^{-4}$. As soon as value of $\eta$ is chosen, the allowed range of $\k$ is $\k>\sqrt{1+\eta}$. Taking into account all this, we chose the following values for the parameters:
\be \label{EtaK}
\eta = 3\times 10^{-4} \,, \qquad \k = \sqrt{1+\eta} + 0.01 \approx 1.01015  \,.
\ee

\begin{figure}
\includegraphics[scale=0.5]{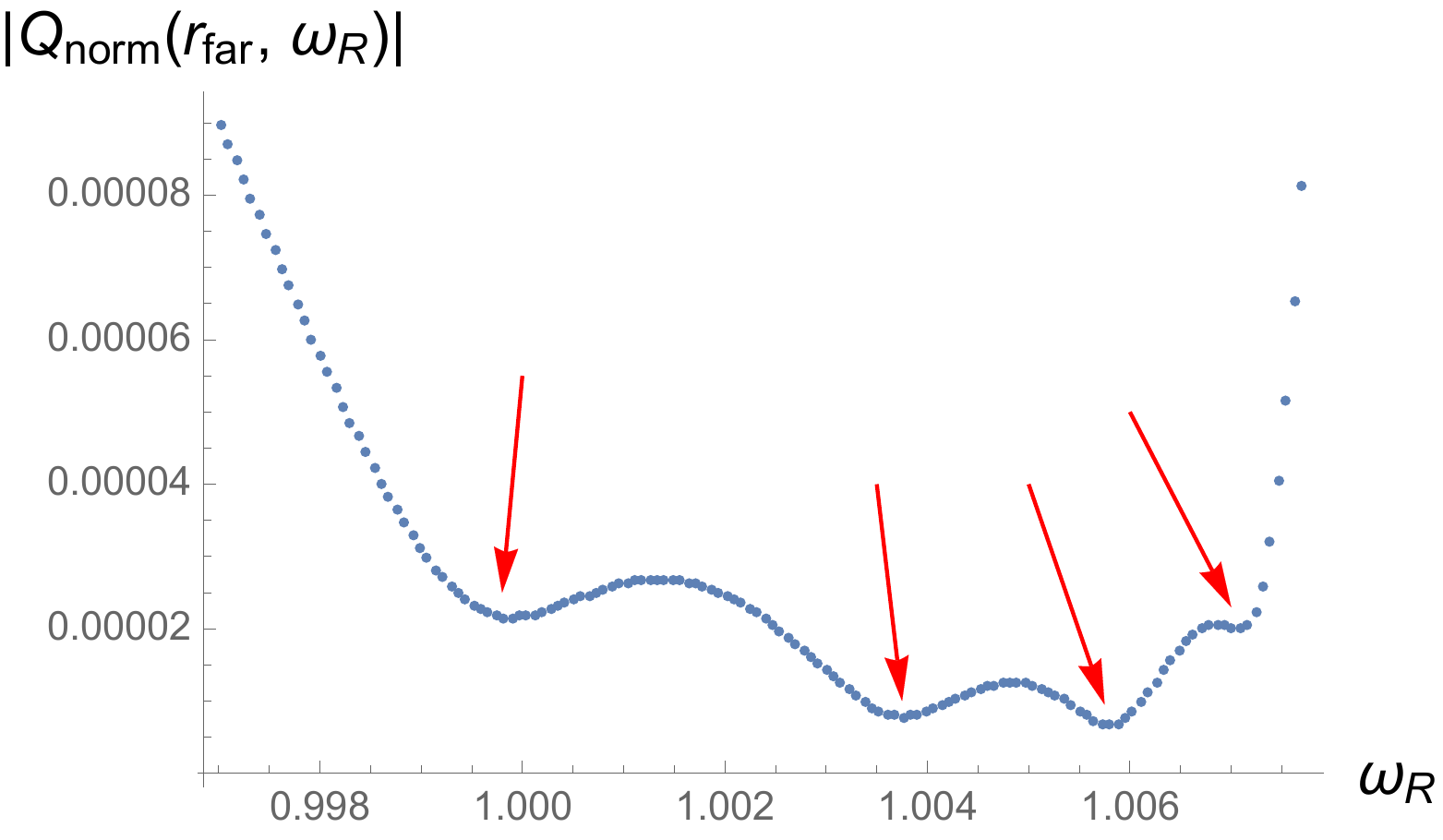}
\caption{\label{Odddipolemins}
Absolute value of $Q_{\rm norm}(r_{far}, \omega_R)$ (defined in Eq.~\eqref{Qnorm}) for $r_{far}=160$, as a function of the real frequency $\omega_R$.  
}
\end{figure}

We follow the procedure described in Sec.~\ref{search_key}. Namely, we determine the near-horizon expansion of $Q$ in Eq.~\eqref{EqQOddDipole}, of the form
\bea
&&Q(r) = (r-r_h)^{-i\omega r_h} \left[ 1 + q_1(r-r_h) + q_2(r-r_h)^2 \right. \nonumber \\
&& \left. + q_3(r-r_h)^3 + q_4(r-r_h)^4 + \cdots  \right] \,.
\eea
Then we numerically integrate Eq.~\eqref{EqQOddDipole}  from $r=1.002$ up to $r_{far} =160$, for a sequence of  real values of $\omega_R$. Then we make a plot of 
\be
|Q_{\rm norm}(r_{far}, \omega_R)| \equiv |Q(r_{far}, \omega_R) e^{-\sqrt{\k^2-\omega_R^2}r_{far}}| \,, \label{Qnorm}
\ee
where the normalizing exponential multiplier is added to compensate the exponential growing behavior that $Q(r, \omega)$ has for all $\omega$'s, except for the frequencies of quasi-bound states. This plot is shown in Fig.~\ref{Odddipolemins}. One can clearly see several minima appearing in this plot. 

\begin{figure}
\includegraphics[scale=0.5]{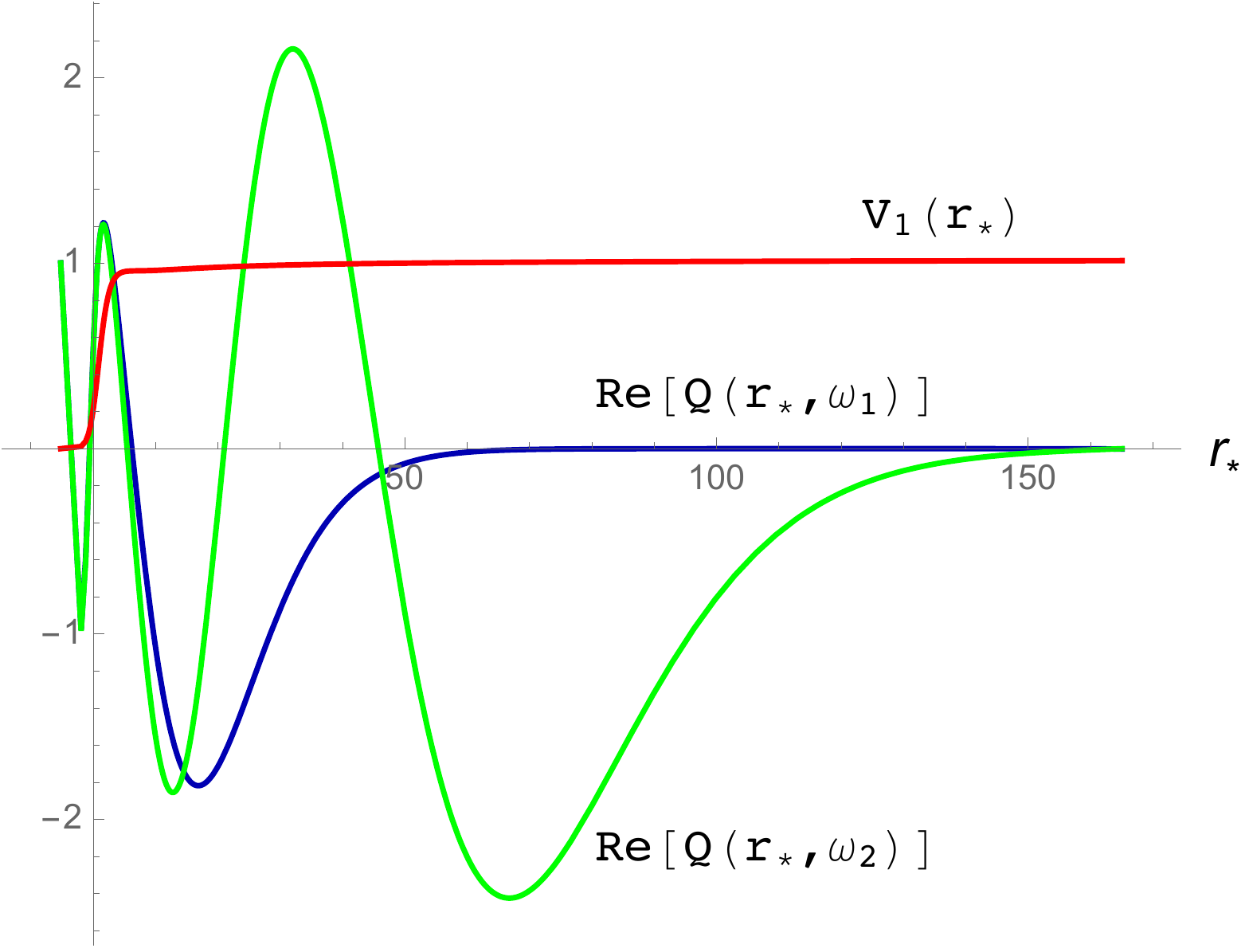}
\caption{\label{Odddipolere}
Real parts of two quasi-bound state solutions $Q(r_{\star})$: the blue curve corresponds to the frequency $\omega_1$ given by \eqref{OddDipOm1}, and the green curve corresponds to the frequency $\omega_2$ given by \eqref{OddDipOm2}. The red curve displays the potential $V_1(r_{\star})$ computed for $r_h=1$ and for $\k, \eta$ given by \eqref{EtaK}.
}
\end{figure}

\begin{figure}
\includegraphics[scale=0.5]{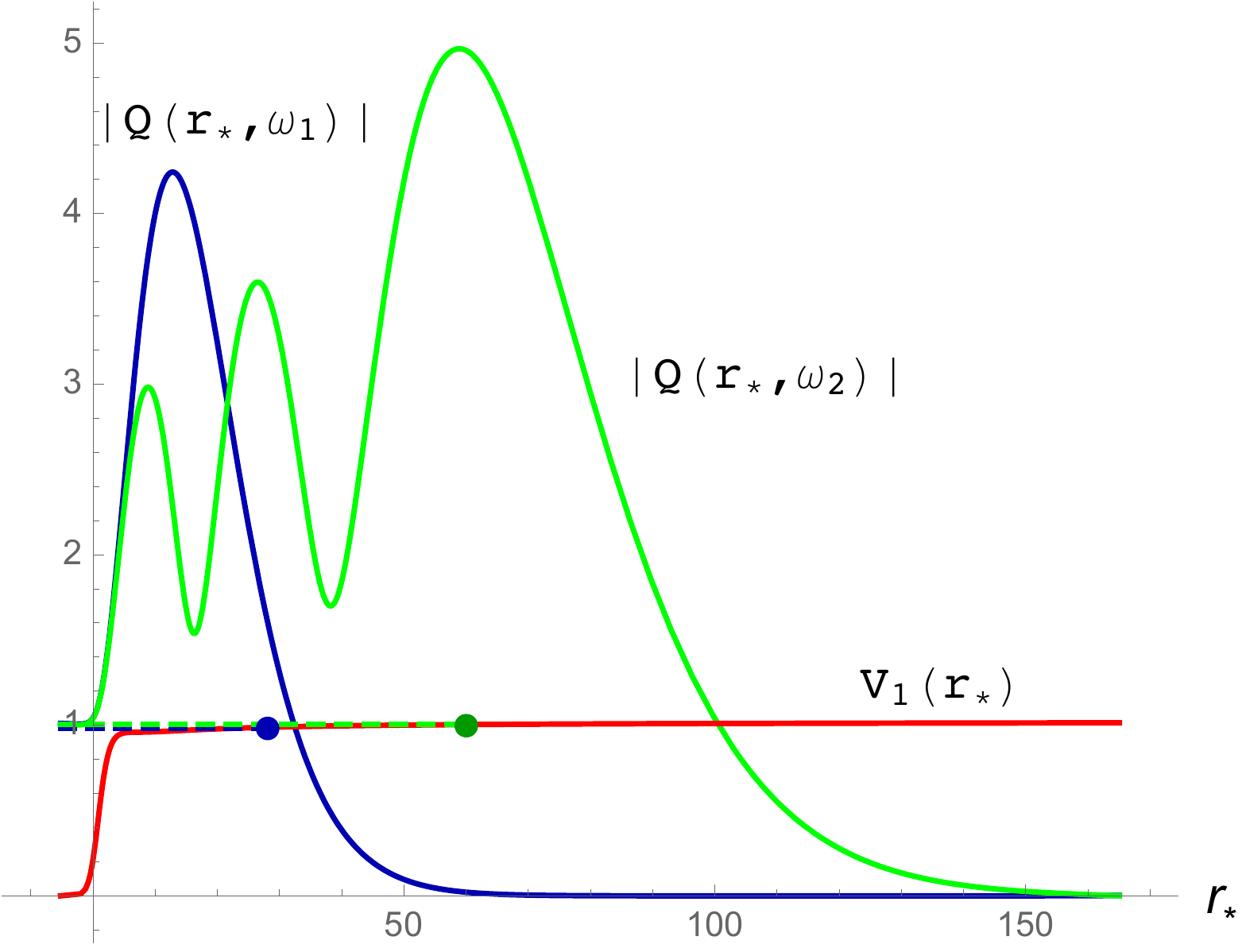}
\caption{\label{Odddipoleabs}
Blue and green solid curves display the absolute values of two quasi-bound state solutions $Q(r_{\star})$: the blue (green) curve corresponds to the frequency $\omega_1$, Eq. \eqref{OddDipOm1} (respectively $\omega_2$,
Eq. \eqref{OddDipOm2}). The red curve displays the potential $V_1(r_{\star})$ computed for $r_h=1$ and for $\k, \eta$ given by \eqref{EtaK}. The blue (green) dashed curve displays the value ${\rm Re}(\omega_{1}^2)$ (${\rm Re}(\omega_{ 2}^2)$). The blue (green) dot marks the radius where the blue (green) dashed line crosses
the  potential. 
}
\end{figure}

The left-most minimum on this plot  (around $\omega_R \approx 1.00$) corresponds to the (quasi) {\it ground} bound state of the odd dipole sector. Applying  Newton's iteration method to obtain the complex value of the frequency of this quasi-bound state, we get
\be \label{OddDipOm1}
\omega^{\rm odd\;dip}_1=0.99050162 - 1.76089 \times 10^{-3} i \,.
\ee
Then we find the solution of the field equation for $\omega=\omega^{\rm odd\;dip}_1$. Corresponding curves, the real part of the function $Q(r, \omega^{\rm odd\;dip}_1)$, and the absolute value of $Q(r, \omega^{\rm odd\;dip}_1)$, are presented in Fig.~\ref{Odddipolere} and Fig.~\ref{Odddipoleabs}, respectively. 

The second minimum (starting from the left)  in Fig.~\ref{Odddipolemins} (around $\omega_R \approx 1.004$) corresponds to a higher harmonic of the odd dipole sector. Applying Newton's iteration method around $\omega_R = 1.004$, we found the frequency of another quasi-bound state,
\be \label{OddDipOm2}
\omega^{\rm odd\;dip}_2=  1.00372252 - 0.49352 \times 10^{-3} i \,.
\ee
The real part  of the solution $Q(r, \omega^{\rm odd\;dip}_2)$ is displayed in Fig.~\ref{Odddipolere}, while its absolute value is displayed in Fig.~\ref{Odddipoleabs}.

Note that both ${\rm Im}[\omega^{\rm odd\;dip}_1]$ and ${\rm Im}[\omega^{\rm odd\;dip}_2]$ are negative, as is necessary in view of the proven linear stability of the odd dipole sector, discussed in the previous subsection.

Let us note that the shapes of the plots of the absolute values of the bound-state field function $Q(r)$ presented on  Fig.~\ref{Odddipoleabs} (including oscillations in the shape of a higher harmonic) is similar to what was found for the bound-states of fermions in a Schwarzschild black hole background \cite{Lasenby:2002mc} (see Figs. 5 and 6 there). 

The Fig.~\ref{Odddipoleenergy} shows the shape of the potential \eqref{PotV1} together with real parts of the  squared frequencies: ${\rm Re}[\omega_1^2]$ (below) and ${\rm Re}[\omega_2^2]$ (above). One can see that quasi-bound states do not correspond to any minimum of the potential. But the potential is flat enough so that 
long-lived quasi-bound states are possible. 

Fig.~\ref{Odddipoleabs} also shows the ${\rm Re}[\omega^2]$ levels of the two studied bound states. Dots mark the radii where the level-lines ${\rm Re}[\omega^2]$ intersect the potential curve $V_1(r)$. Looking at the position of these two points, one can see that the exponential decay of the solutions indeed starts more or less where the energy becomes smaller than the potential, ${\rm Re}[\omega^2]<V_1(r)$. This is more evident for the first quasi-bound state than for the second one. The reason is probably that, for the second quasi-bound state, the energy line intersects the potential in a region where the potential is very flat, so that the wave can ``tunnel'' more easily below the potential.

\begin{figure}
\includegraphics[scale=0.5]{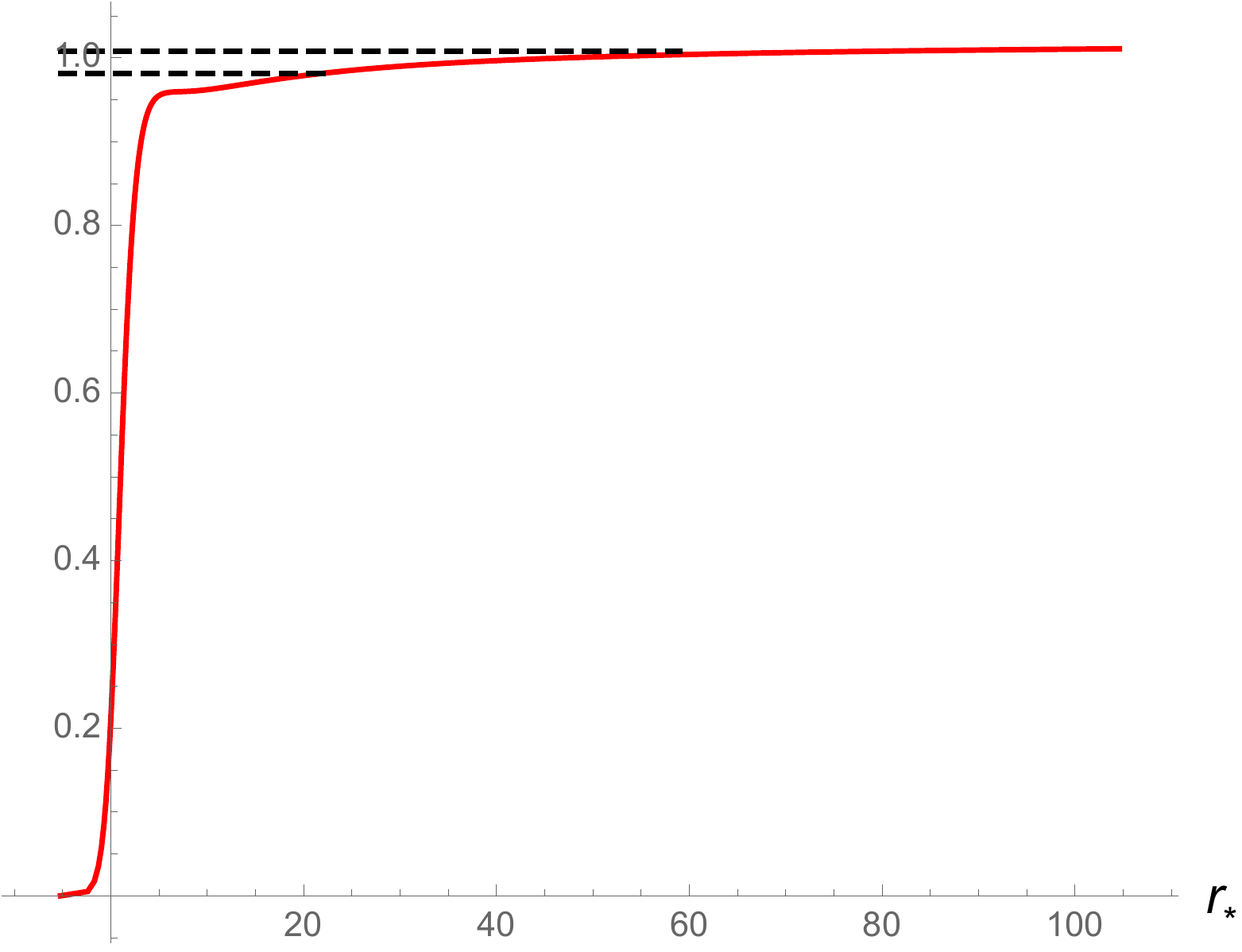}
\caption{\label{Odddipoleenergy}
The red solid curve displays the potential $V_1(r_{\star})$. The two dashed lines display the squared frequencies 
${\rm Re}(\omega_1^2)$ and ${\rm Re}(\omega_2^2)$ (Eqs.~\eqref{OddDipOm1} and \eqref{OddDipOm2}, respectively).
}
\end{figure}

\section{Quasi-bound states and stability of the even dipole sector}\label{EvenDipStab}
The dipole perturbations of the even sector are described by two second-order equations. As far as we know,
 the stability can not be established directly as it was done in the case of one second-order equation describing the odd dipole perturbations. Thus, we concentrate on searching for quasi-bound states. As we discuss now,
 all the quasi-bound states we  found have negative imaginary parts, and this is an evidence for stability. 

Before proceeding, let us mention the near-horizon behavior of all the variables of the even sector \eqref{6varNewSec}. Using the Lorentz-transformation properties of the frame
components $u_{ij}$ of the tensor $u_{\mu\nu}$, Eq. \eqref{eqGrav}, from a (singular) \sch-coordinate related 
frame to a regular-coordinate related frame, one obtains the following regularity properties, near the horizon,
for the seven relevant coordinate components of $u_{\mu\nu}$:
\bea \label{evenregularity}
&&\overline{h}_0 \sim u^{\rm even}_{02} \sim (r-r_h)^{0}  \,, \nonumber \\
&&\overline{h}_1 \sim u^{\rm even}_{12} \sim (r-r_h)^{-1} \,, \nonumber  \\
&&H_0 \sim u^{\rm even}_{00}/f(r) \sim (r-r_h)^{-1}  \,, \nonumber \\
&&H_1 \sim u^{\rm even}_{01} \sim (r-r_h)^{-1} \,, \nonumber \\
&&H_2 \sim u^{\rm even}_{11} f(r) \sim (r-r_h)^{-1} \,, \nonumber \\
&&G \sim \frac{1}{2}(u^{\rm even}_{22}\sin^2{\theta}-u^{\rm even}_{33}) \sim (r-r_h)^{0}   \,, \nonumber \\
&&K \sim \frac{1}{2}(u^{\rm even}_{22}\sin^2{\theta}+u^{\rm even}_{33}) \sim (r-r_h)^{0} \,.
\eea
Here all the indices of the tensor $u_{\mu\nu}$ are space-time indices in the $(t,r,\theta, \phi)$ \sch coordinate system.

Now consider the system of equations \eqref{EqEvenL1} which describes the even dipolar perturbations. Taking into account the regularity properties \eqref{evenregularity}, and substituting in the system \eqref{EqEvenL1} an ansatz of the type \eqref{NHE}, we eventually obtain the following 
near-horizon expansion for the variables $\overline{h}_0, \overline{h}_1, H_1, K$:
\begin{widetext}
\bea
&& \overline{h}_0=(r-r_h)^{-i\omega r_h}\left[  h_{0(0)}(p_h,p_H) +h_{0(1)}(p_h,p_H)(r-r_h)+h_{0(2)}(p_h,p_H)(r-r_h)^2+ \cdots \right] \,, \nonumber \\
&& K=(r-r_h)^{-i\omega r_h}\left[  K_{(0)}(p_h,p_H)+K_{(1)}(p_h,p_H)(r-r_h)+K_{(2)}(p_h,p_H)(r-r_h)^2+ \cdots \right] \,, \nonumber \\
&& \overline{h}_1=\frac{(r-r_h)^{-i\omega r_h}}{r-r_h}\left[  p_h+h_{1(1)}(p_h,p_H)(r-r_h)+h_{1(2)}(p_h,p_H)(r-r_h)^2+ \cdots \right] \,, \nonumber \\
&& H_1=\frac{(r-r_h)^{-i\omega r_h}}{r-r_h}\left[  p_H+H_{1(1)}(p_h,p_H)(r-r_h)+H_{1(2)}(p_h,p_H)(r-r_h)^2 + \cdots \right] \,. \quad \label{icondevenL1}
\eea
\end{widetext}
This near-horizon expansion depends on two free parameters, say, $p_h,p_H$. 

To explore some concrete quasi-bound states in the even dipole sector, we choose the same values of $\eta$ and $\k$ as those we used in the odd dipole case, i.e.,
\be
\eta = 3\times 10^{-4} \,, \qquad \k = \sqrt{1+\eta} + 0.01 \approx 1.01015  \,.
\ee
Then we choose a set of (real) values $\omega_R \lesssim \k$, and, for each value of $\omega_R$ in this set, we integrate the system \eqref{EqEvenL1} twice, once with the initial conditions \eqref{icondevenL1} with $p_h=1,\,p_H=0$, and once with those with $p_h=0,\,p_H=1$. The starting point of integration is $r=1.0001$, and the final point is $r_{far}=150$. Then, for each value of $\omega_R$, we compute the determinant
\be  \label{evendet}
{\rm d}(\omega_R) \equiv \det{\begin{pmatrix} \overline{h}_{0}(r_{far})|_{p_h=1,p_H=0} & \overline{h}_{0}(r_{far})|_{p_h=0,p_H=1} \\  K(r_{far})|_{p_h=1,p_H=0} & K(r_{far})|_{p_h=0,p_H=1} \end{pmatrix}} \,,
\ee
and we plot the absolute value of ${\rm d}(\omega_R)$, normalized for each column, by the exponential factor already known from the Sec.~\ref{OddSec}, i.e., 
\be \label{RenormDet}
|{\rm d}(\omega_R)| e^{-2\sqrt{\k^2-\omega_R^2}r_{far}} \,.
\ee 

\begin{figure}
\includegraphics[scale=0.6]{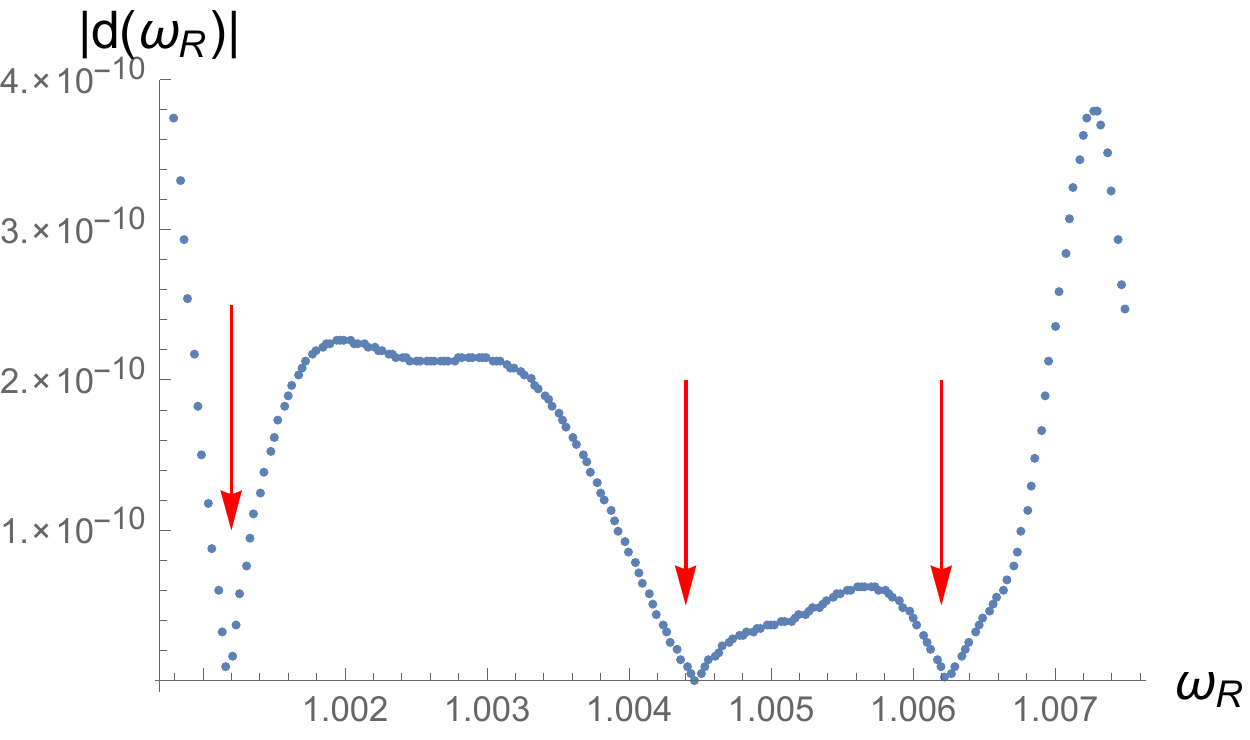}
\caption{\label{evendipmins}
Absolute value of the renormalized determinant ${\rm d}(\omega_R, r_{far})$ (defined in \eqref{RenormDet}) for $r_{far}=150$, as a function of the real $\omega_R$. 
}
\end{figure}

The corresponding plot is shown in Fig.~\ref{evendipmins}. We again see a few clear minima of the absolute value of the normalized determinant \eqref{RenormDet}. We computed two quasi-bound states corresponding to the first two minima on the plot (counted from the left). After using Newton's iteration method to find complex solutions $\omega^{\rm even\; dip}$ of the complex equation 
\be
{\rm d}(\omega^{\rm even\; dip})=0 
\ee
(each time with starting points near the corresponding minimum), we found the following complex values of the quasi-bound state frequencies,
\bea
&& \omega_1^{\rm even\; dip}= 1.001177699477 - 1.39409 \times 10^{-8} i \,,\quad \\
&& \omega_2^{\rm even\; dip}= 1.004461851066 - 1.10375 \times 10^{-8} i \,.\quad
\eea
The imaginary parts of both complex frequencies are negative. Studying other bound states with higher frequencies, we also found each time the expected (negative) sign of the imaginary part.

\section{Quasi-bound states and stability of the even sector for $L\geq 2$} \label{EvenMulti}
In this section, we study quasi-bound states in the even sector of perturbations for $L \geq 2$. To work with concrete numerical values, we consider $L=2$. [This choice is based on: (i) the formal link ($\eta \to -1$) between torsion bigravity perturbations and bimetric gravity ones; (ii) the stability of $L \geq 1$ bimetric gravity perturbations \cite{Brito:2013wya}; and (iii)  the intuition that larger angular momentum values are less prone to instabilities.]
 As to values of $\k$ and $\eta$, we take (in units $r_h=1$) $\k=2$ and $\eta=1$, in order to enlarge the domain of parameters, compared to what was used in Sec.~\ref{OddSec} and \ref{EvenDipStab}. 

In Sec.~\ref{evendipsec} we explained how the initial system of equations describing the multipolar even sector transforms into a system \eqref{sysvar6},
\be
V_6^{'}(r)=A_6[r, r_h,\k,\eta,L,\omega]V_6(r) \;,  \label{sysvar6New}
\ee
of 6 first-order equations for the 6 variables 
\be
\{ G, \overline{h}_0, H_0, H_1, H_2, K \} \;. \label{6varNewSec}
\ee

The system \eqref{sysvar6New}, when considered near the horizon, is singular, because of the presence of $(r-r_h)^{-3}$ denominators in the matrix $A_6(r)$. In addition, it is singular at infinity, because of a power-law growth of $A_6(r)$ as $r \to \infty$.   Thus, this system is not convenient either for expressing near-horizon boundary conditions or for integration. But one can construct two different systems of three {\it second-order} ODEs, such that each of them has 
a nice behavior either on the horizon, or at infinity (though not at both!). 

First, one can transform the system \eqref{sysvar6New} into a second-order system
\be
V_{{\rm inf}\;3}^{''} = M_{H_0H_1K} V_{{\rm inf}\;3} \label{sys01K}
\ee
for the three variables $V_{{\rm inf}\;3} \equiv H_0, H_1, K$. This system has the following limit at $r\to\infty$:
\be
V_{{\rm inf}\;3}^{''}=(\k^2-\omega^2)V_{{\rm inf}\;3} + O\l \frac{1}{r}\r V_{{\rm inf}\;3} + O\l \frac{1}{r}\r V^{'}_{{\rm inf}\;3} \;, \label{V3As}
\ee
as it supposed to be in the far wave zone for a massive particle with mass $\k$.
Let us note in passing that, taking into account the additional terms of order $1/r$ in Eq.~\eqref{V3As}, we obtained the following asymptotic behavior of $H_0, H_1,K$ near infinity:
\bea
&&K_{\rm As} \sim e^{-\sqrt{\k^2-\omega^2}r}r^{-1+r_h(2\omega^2-\k^2)/(2\sqrt{\k^2-\omega^2})} \;, \nonumber \\
&&H_{0\;{\rm As}} = \frac{2(\k^2-\omega^2)}{\k^2}K_{\rm As} \;, \nonumber \\
&&H_{1\;{\rm As}} = \frac{2i\omega\sqrt{\k^2-\omega^2}}{\k^2}K_{\rm As} \;.
\eea
However, while having a good behavior near infinity, this system is singular at the horizon.

A horizon-regular second-order system can be obtained from the six first-order equations \eqref{sysvar6New} by working with the three variables $K, G$ and $H_{20} \equiv H_2-H_0$. [This can be seen from the definition of $H_0$ and $H_2$, and from the Lorentz transformation laws of the corresponding components of $u_{\mu\nu}$.] When considered at $r\to r_h$, this system takes the following form:
\be  \label{syshor}
\frac{d}{d r_{\star}} V_{{\rm hor}\;3}  \approx -\omega^2 V_{{\rm hor}\;3} \;.
\ee
On the other hand, this system is singular at $r \to \infty$. 

To look for quasi-bound states, we use the following strategy. 
First, we need to set initial conditions near the horizon. To this end, we insert in the horizon-regular system \eqref{syshor}\footnote{By contrast, it turned out to be quite tricky to obtain the near-horizon initial conditions by using the system \eqref{sys01K}.} the following ansatz:
\bea \label{GKHnearhor}
&&  G|_{r\to r_h} = (r-r_h)^{-i \omega r_h}\left[   a_{g\,0} + \Sigma^{n}_{i=1}a_{g\,i}(r-r_h)^i \right] \;, \nonumber \\
&& K|_{r\to r_h} = (r-r_h)^{-i \omega r_h}\left[   a_{k\,0} + \Sigma^{n}_{i=1}a_{k\,i}(r-r_h)^i \right] \;,  \nonumber \\
&&  H_{20}|_{r\to r_h} = (r-r_h)^{-i \omega r_h}\left[   a_{h\,0} + \Sigma^{n}_{i=1}a_{h\,i}(r-r_h)^i \right] \;. \nonumber \\
\eea
Here the order $n$ is related to the precision of the initial conditions; for our calculations, we used $n=5$. The power factor in front selects only ingoing waves, see the Sec.~\ref{search_key}. Then, solving the system \eqref{syshor} with such an ansatz, we obtain all the $a_{g\,i}, a_{k\,i}, a_{h\,i}$ expressed through the three free parameters $a_{g\,0}, a_{k\,0}, a_{h\,0}$. Thus, we determined the near-horizon expansion \eqref{GKHnearhor} for $G,K,H_{20}$.

Then, using the relations between the variables $\{H_0, H_1\}$ and $\{G,K, H_{20}\}$ [obtained as a by-product of transforming the system \eqref{sysvar6New} to the system \eqref{syshor}] and the near-horizon expansion \eqref{GKHnearhor}, we could obtain the near-horizon expansions for the variables $H_0$ and $H_1$. This finally gave us near-horizon expansions for the three variables $H_0, H_1, K$ entering the infinity-regular system \eqref{sys01K}, namely,
\bea \label{01kbound}
H_0|_{r \to r_h} &=& H_0^{\rm boundary}(a_{g\,0}, a_{k\,0}, a_{h\,0}) \;, \nonumber \\
H_1|_{r \to r_h} &=& H_1^{\rm boundary}(a_{g\,0}, a_{k\,0}, a_{h\,0}) \;, \nonumber \\
K|_{r \to r_h} &=& K^{\rm boundary}(a_{g\,0}, a_{k\,0}, a_{h\,0}) \;.
\eea

As a second step in our strategy, we integrate the infinity-regular system \eqref{sys01K}, starting from horizon, with the near-horizon boundary conditions \eqref{01kbound}. As explained before, this integration is repeated three times, with the following three sets of initial conditions,  $a_{g\,0}=1, a_{k\,0}=0, a_{h\,0}=0$, for $a_{g\,0}=0, a_{k\,0}=1, a_{h\,0}=0$ and for $a_{g\,0}=0, a_{k\,0}=0, a_{h\,0}=1$. This constructs three ``basis" solutions, similarly to what was described in Sec.~\ref{search_key} and used in Sec.~\ref{EvenDipStab}.

For a sequence of real values of $\omega$, we integrate the system \eqref{sys01K}.
We compute the determinant of the three ``basis" solutions, at a radius $r_{far} \gg r_h$  (``at infinity"). We study the absolute value of the determinant\footnote{Renormalized by a factor $e^{-3\sqrt{\k^2-\omega^2}r_{far}}$.} as a function of (real) $\omega$, looking for minima. Once a minimum is located, we search for complex values of the quasi-bound state frequencies using Newton's iteration method.

Using this procedure, we found the following two quasi-bound states:
\be
\omega^{{\rm even}}_1 =  1.96222617   - 1.22780 \times 10^{-3} i \;,
\ee
and
\be
\omega^{{\rm even}}_2 = 1.97941109 - 0.80466 \times 10^{-3} i \;.
\ee
In both cases the imaginary part of frequency is negative, which gives evidence for stability.

\section{Superradiance instability around rotating black holes in torsion bigravity} \label{super}
Most of the discussions in the literature of superradiance instabilities around Kerr black holes (e.g., \cite{Brito:2020lup, Stott:2020gjj}) focus on the 
small-mass case, $ \k r_ h <1$. However, because of the constraint \eqref{c}, we are interested in the opposite case, $ \k r_ h >1$. 

It is useful to start our discussion by using approximate but explicit analytical descriptions of superradiance instabilities around rotating black holes.
Zouros and Eardley \cite{Zouros:1979iw} gave an analytic estimate of the characteristic time of development of superradiant instabilities in the case of a massive {\it scalar} field around a \sch black hole. Their calculations concerned
the case $ \k r_ h >1$ (which is the one relevant for us),
and were based on the WKB approximation (applicable when $ \k r_ h \gg 1$). Their result (for
fast-spinning black holes) was:
\bea \label{ZE}
\tau_{ZE} &\sim& 10^7 e^{1.84 M\k}\l \frac{G M}{c^3} \r  \nonumber \\
&=& 50\; e^{1.84 \frac{M}{M_{\odot}}\; M_{\odot}\k}\l \frac{M}{M_{\odot}} \r \;{\rm sec} \;,
\eea
where $M$ is the mass of the black hole. It is reasonable to use the Salpeter time\footnote{Which is a characteristic time of a black hole growth.} as a reference timescale, 
\be
\tau_S = 4.6\times 10^7 \;{\rm yr} \approx 1.4\times 10^{15} \; {\rm sec}  \;,
\ee
to be compared with the timescale of instability. Thus, for each detected spinning black hole of mass $M$, the values of $\k$ which do not contradict the existence of this spinning black hole must be such that $\tau_{\rm inst}(\k, M) \gtrsim \tau_{\rm S}$. This gives the following lower bound on $\k$:
\be
\k \; \gtrsim \; 0.54 \l\frac{M}{M_{\odot}}\r^{-1} \log{\left[ 2.8\times 10^{13}  \l\frac{M}{M_{\odot}}\r^{-1} \right]}  \frac{c^2}{G M_{\odot}}  \;.
\ee
For example, if one would observe a (fast) spinning black hole of mass $M \sim M_{\odot}$, this would restrict the value  of the mass (or inverse range) of the massive field to satisfy
$\k \gtrsim (90 \;{\rm m})^{-1}$. But we do not have evidence for the existence of such a black hole.
The observation of the most studied highly spinning black hole candidate Cygnus X-1, whose mass is about $21 M_{\odot}$, gives the following restriction:
\be
\k \gtrsim (2 \; {\rm km})^{-1} \approx 10^{-10}\;{\rm eV}\; ({\rm from \; Eq.} \eqref{ZE}).
\ee
However, this is  an approximate WKB-based estimate, and it only concerns a massive {\it scalar} field.
A more accurate, and directly relevant\footnote{As we are generically interested in situations where $\k r_h \gg1$,
the additional Weyl-curvature coupling in Eq. \eqref{eqTor} is expected to have a negligible effect, so that the
Fierz-Pauli-based computation of \cite{Brito:2020lup} is applicable to torsion bigravity.}, computation was made by Brito, Grillo and Pani in Ref.~\cite{Brito:2020lup} (see also \cite{Stott:2020gjj}). Their analysis was made for a massive {\it tensor} field around a \sch black hole, and was not relying on the WKB approximation. They numerically computed the characteristic time of instability (see Eq.~(6) there), which depends on the mass of the black hole, its spin, and the mass of the spin-2 field. Comparing this characteristic time with the Salpeter time, Ref.~\cite{Brito:2020lup} obtains, for each particular value of $\k$, the region in the mass-spin plane where $\tau_{\rm inst}(\k) < \tau_{\rm S}$. The mass-spin diagram Fig.~2 of \cite{Brito:2020lup} shows the mass-spin regions for which $\tau_{\rm inst}(\k) < \tau_{\rm S}$, for various values of $\k$ ($\mu$ in their notation). In particular, the observation of Cygnus X-1 with its values of spin and mass forbids the values $ 10^{-13} \lesssim \k/{\rm eV} \lesssim 10^{-11}$. For the case of torsion bigravity where $\k$ is anyway bounded from below, $\k > \sqrt{1+\eta}/r_h$, we conclude that, in order not to have observable effects of superradiance instabilities, we should impose\footnote{Brito, Grillo and Pani \cite{Brito:2020lup} obtained  several other accurate bounds for $\k$. In particular, they compared the characteristic time of instability against the baseline during which the spin of Cygnus X-1 was measured to be constant. Another, indirect, bound was obtained by using the electromagnetic estimates of black hole spins obtained from Ka iron line data. However, all the bounds give the same result in order of magnitude, which is sufficiently accurate for the schematic consideration we are carrying on here.}
\be
\k \gtrsim 10^{-11} \; {\rm eV} \approx  (20 \; {\rm km})^{-1} \;.
\ee
 As already mentioned in the Introduction, this does not put any new limits on torsion bigravity, compared to 
 the 
 limit $\k > \sqrt{1+\eta} \,(6 \; {\rm km})^{-1}  $ obtained in \cite{Nikiforova:2021xcj} [see Eq.~(7.1) there], from the condition of not having singularities in spherically-symmetric perturbations (assuming the existence of a $2 M_{\odot}$ black hole). The factor $ \sqrt{1+\eta}$ can be approximated by 1, when taking
 into account the phenomenological limit $\eta \lesssim 3\times 10^{-4}$ obtained in Ref.~\cite{Damour:2019oru} for the kilometer-range case.

\section{Conclusions}
We investigated the non spherically-symmetric massive spin-2 perturbations around a \sch black hole in torsion bigravity.  We gave strong evidence for stability of perturbations, and exhibited some quasi-bound states. 

In future work, we intend to complete the present results by studying the vibration modes of a \sch black hole in torsion bigravity, i.e., the quasi-normal modes. First one must investigate the spectrum of  quasi-normal modes of the massive spin-2 perturbations, and then investigate the way they influence the massless spin-2 field $h_{\mu\nu}$ [According to the triangular structure of the perturbed field equations which is illustrated in Eqs.~\eqref{EqEinstEf}--\eqref{TmunuEf}, the linear massless spin-2 perturbations decompose as the sum of usual GR perturbations plus
contributions sourced
by the massive spin-2
perturbations]. The perturbations $h_{\mu\nu}$ of the metric are directly observable  by LIGO-Virgo gravitational wave detectors, which makes this study of physical interest. 
 
\section*{Acknowledgments}
The author thanks Thibault Damour and Fidel Schaposnik  for useful suggestions.

\newpage
\appendix

\section{} \label{apA}
Denoting $W_L(\theta)=\d^2_{\theta}P_L - \cot{\theta}\d_{\theta}P_L $ (where $P_L(\cos{\theta})$ is the usual Legendre polynomial), we decompose $u_{\mu\nu}$
according to (using $M=0$ to simplify the computations)

\begin{widetext}
\be
u_{\mu\nu}^{{\rm odd}}(t,r,\theta,\phi)=\begin{pmatrix} 
0\;&\;0\;&\;0 \;&\; -h_0(t,r)\,\sin{\theta}\, \d_{\theta}P_L (\cos{\theta}) \\ 
\\
\star\;&\;0\;&\;0 \;&\; -h_1(t,r)\,\sin{\theta}\, \d_{\theta}P_L (\cos{\theta}) \\ \\
\star & \star & 0 & h_2(t,r) \sin{\theta} W_{L}(\cos{\theta}) \\ \\

\star\;&\; \star\;&\;\star\;&\; 0
   \end{pmatrix} \,, \label{DecOdd}
\ee

\be
u_{\mu\nu}^{{\rm even}}(t,r,\theta,\phi)=\begin{pmatrix} 
f(r)H_0(t,r) P_L \;&\;H_1(t,r) P_L \;&\;\overline{h}_0(t,r)\d_{\theta}P_L  \;&\; 0  \\ 
\\
\star\;&\;f(r)^{-1}H_2(t,r) P_L \;&\;\overline{h}_1(t,r)\, \d_{\theta}P_L  \;&\; 0  \\ \\
\star \;&\; \star \;&\; r^2\left[  K(t,r) P_L  + G(t,r) W_{L}  \right] \;&\; 0 \\ \\

\star\;&\; \star\;&\;\star\;&\; r^2 \sin^2{\theta} \left[  K(t,r) P_L  - G(t,r) W_{L}  \right]
   \end{pmatrix} \,. \label{DecEven}
\ee
\end{widetext}






\end{document}